# ARTICLE

# Dielectric metasurface-assisted cavity ring-down spectroscopy for thin-film circular dichroism analysis

Ankit Kumar Singh,[†,a] Zhan-Hong Lin,[†,a] Min Jiang,[a] Thomas G. Mayerhöfer,[a,b] and Jer-Shing Huang*[,a,b,c,d,e]

Chiral molecules show differences in their chemical and optical properties due to different spatial arrangements of the atoms in the two enantiomers. A common way to optically differentiate them is to detect the disparity in the absorption of light by the two enantiomers, i.e. the absorption circular dichroism (CD). However, the CD of typical molecules is very small, limiting the sensitivity of chiroptical analysis based on CD. Cavity ring-down spectroscopy (CRDS) is a well-known ultrasensitive absorption spectroscopic method for low-absorbing gas-phase samples because the multiple reflections of light in the cavity greatly increase the absorption path. By inserting a prism into the cavity, the optical mode undergoes total internal reflection (TIR) at the prism surface and the evanescent wave (EW) enables the absorption detection of condensed-phase samples within a very thin layer near the prism surface, called EW-CRDS. Here, we propose an ultrasensitive chiral absorption spectroscopy platform using a dielectric metasurface-assisted EW-CRDS. We theoretically show that, upon linearly polarized and oblique incidence, the metasurface exhibits minimum scattering and absorption loss, introduces negligible polarization change, and locally converts the linearly polarized light into near fields with finite optical chirality, enabling CD detection with EW-CRDS that only works with linearly polarized light. We evaluate the ring-down time in the presence of chiral molecules and determine the sensitivity of the cavity as a function of total absorption from the molecules. The findings open the avenue for an ultrasensitive thin film detection of the chiral molecules using the CRDS techniques.

## Introduction

Chirality is the property of a three-dimensional object which makes the superposition of the object and its mirror image impossible for any translation or rotation. The chirality of molecules influences many biological processes because of the chiral selection of the biological entities, which are made from an inherently chiral amino acid [1,2]. Therefore, detecting the chirality of molecules and their separation is a topic of immense scientific research importance [3-6]. The identical physicochemical properties of the enantiomers make it difficult to economically distinguish them in a physical and chemical system. Circular dichroism (CD), defined as differential absorption of the two-handedness of circularly polarized light, serves as the most convenient and economical way to distinguish molecular enantiomers [7,8]. However, CD response is very weak, i.e., typically $10^{-5}$ times the average dipolar absorption, and thus not sensitive enough for quantitative analysis at low concentrations. The CD response of a molecule is proportional to the optical chirality (OC) given as $2\omega\, imag(\boldsymbol{E}^*.\boldsymbol{B})/\epsilon$, where $\boldsymbol{E}$ ($\boldsymbol{B}$) is electric (magnetic) field, $\omega$ is the angular frequency and $\epsilon$ is the permittivity of the medium [9,10]. Therefore, to enhance the absorption difference, one seeks methods to provide an optical field with parallel $\boldsymbol{E}$ and $\boldsymbol{B}$ field components with a $\pi/2$ phase difference. Various plasmonic and dielectric nanostructures were proposed to enhance differential chiral absorption in the near field by enhancing the OC of the field [11-16]. These nanostructures include both chiral and achiral structures designed for excitation with linear and circular polarizations.

Apart from enhancing the CD by optical field engineering, one must also consider the Beer–Lambert law for evaluating the concentration of the chiral species in a real device. Based on the Beer–Lambert approximation [17], $A = \varepsilon bC$, where $A$ is the absorbance, $\varepsilon$ is molar absorptivity, $b$ is the absorption path length and $C$ is the concentration. The CD signal, i.e. the difference in the absorption, can be expressed as $\Delta A = (\varepsilon_R - \varepsilon_L)bC$ [18]. Here, $\varepsilon_R$ ($\varepsilon_L$) represents the molar absorptivity of right (left) hand molecules. To enhance the sensitivity of CD measurement, i.e. $\Delta A/C$, it is necessary to increase $(\varepsilon_R - \varepsilon_L)$ and $b$. While $(\varepsilon_R - \varepsilon_L)$ is related to the chiral absorption from the molecule and hence directly proportional to the OC of the optical field, $b$ can be enhanced by using cavity ring-down spectroscopy (CRDS), which is a well-established method to improve weak absorption signal by enhancing absorption length

[a.] *Leibniz Institute of Photonic Technology, Member of Leibniz Health Technologies, Member of the Leibniz Centre for Photonics in Infection Research (LPI), Albert Einstein Straße 9, 07745 Jena, Germany.*
[b.] *Institute of Physical Chemistry, Friedrich Schiller University Jena, Member of the Leibniz Centre for Photonics in Infection Research (LPI), Helmholtzweg 4, 07743 Jena, Germany.*
[c.] *Abbe Center of Photonics, Friedrich Schiller University Jena, Member of the Leibniz Centre for Photonics in Infection Research (LPI), Helmholtzweg 4, 07743 Jena, Germany.*
[d.] *Research Center for Applied Sciences, Academia Sinica, Taipei 11529, Taiwan.*
[e.] *Department of Electrophysics, National Yang Ming Chiao Tung University, Hsinchu 30010, Taiwan*
† Equal contribution.

$b$ through multiple reflections of the light inside an optical cavity [19,20]. The "ring-down time ($\tau$)", i.e., the decay time of the cavity mode, can be used to estimate the absorption coefficient of a weakly absorbing analyte. For condensed-phase samples, evanescent-wave cavity ring-down spectroscopy (EW-CRDS) has been demonstrated as an extremely powerful tool for measuring the absorption from weakly absorbing samples [21]. However, using EW-CRDS for CD measurement is not simple because circularly polarized light cannot survive in the EW-CRDS cavity due to the differential reflectance of s- and p- polarized components at the dielectric interfaces. Since only linearly polarized light is suitable for EW-CRDS, Sofikitis et al. [22] have set up bowtie ring cavities with Faraday rotators to measure circular birefringence, which rotates the polarization of the linearly polarized light due to the molecular chirality-dependent real part of the refractive index of the chiral molecules. With the help of the Kramers-Kronig relations, the measured spectrum of the real index can be used to calculate the spectrum of the imaginary index that links to the CD. This method requires, however, the birefringence spectrum to be recorded over a certain spectral range. This is experimentally demanding because the high reflectivity of the mirrors and the Faraday rotator used in CRDS is typically optimized for a narrow bandwidth and obtaining a broadband performance from both the components is difficult.

Here, we propose using a dielectric metasurface to assist EW-CRDS for direct CD measurement of thin chiral film or low-absorption chiral samples at small concentrations. Upon excitation of the evanescent field of the totally internally reflected linearly polarized cavity mode, the metasurface converts the excitation field into optical near fields with a finite OC, enabling direct CD measurement with linearly polarized excitation in the EW-CRDS cavity.

## Results and discussion

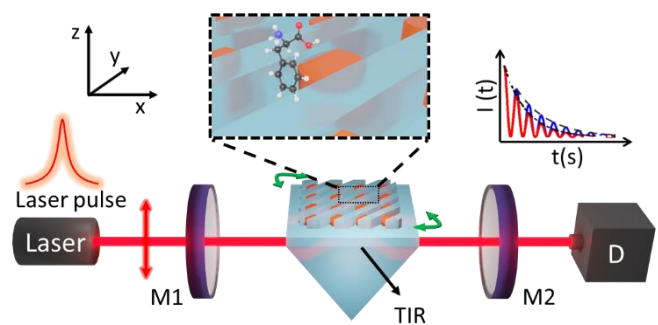

**Figure 1:** The proposed experimental scheme for the detection of chiral molecules using EW-CRDS. A linearly polarized laser pulse is introduced in the cavity consisting of highly reflective mirrors (MI and M2) and a prism designed such that the light is incident at Brewster's angle on the side faces. The metasurface is placed on the top of a prism with a thin layer of index-matching liquid to flexibly move and rotate the metasurface (indicated by the green curved double arrows). The time-dependent intensity response of the light emitted from the cavity is recorded using a photomultiplier tube (D).

Figure 1 shows the proposed scheme for CD analysis of chiral samples at low concentrations using dielectric metasurface-assisted EW-CRDS. The optical cavity is a linear cavity consisting of two highly reflection (> 0.99995%) mirrors and a specially designed glass prism with a refractive index of 1.5. Either a pulsed or chopped CW laser can be used to excite the cavity [19]. The prism is designed such that p- polarized light along the axis of the linear cavity is incident at Brewster's angle of the air/glass interface and undergoes total internal reflection (TIR) at the top surface of the prism (Figure 1). Brewster's angle incident helps minimize the loss since p-polarized incidence at Brewster's angle can theoretically reach a zero reflection at the interface without using any anti-reflection coating. The scheme of EW-CRDS is indeed well-established for measuring the absorption from thin films and other low-absorption condensed phase samples. However, directly applying a conventional EW-CRDS cavity for CD measurement is not possible because circularly polarized light cannot survive in the cavity.

To make the cavity suitable for CD measurement, we propose using a dielectric metasurface to convert the linearly polarized evanescent field of the cavity mode into a chiroptical active near field, i.e., a near field with a non-zero OC. The metasurface on top of a cover glass and immersed in water is designed to be placed on top of the TIR surface of the prism. The metasurface is a simple two-dimensional array of gallium phosphide (GaP) nanopillars of radius 80 nm and height 50 nm arranged with a period of 260 nm in x and y directions. Since linearly polarized excitation leads to a near-field OC distribution around the pillars with alternating handedness, it is necessary to selectively block the area with unwanted handedness [13]. Otherwise, the OC and thus the CD would be averaged out to zero. For this, diagonally orientated glass stripes (height = 160 nm and width (T) = 70 nm) are designed as partial protectors to block the area with unwanted near-field OC (Figure 2a). In this way, the sample can only access near-field OC with one selected handedness, which can be easily switched by in-plane rotation of the metasurface, as illustrated by the green double arrow in Figure 1. Therefore, this method is able to obtain the CD of chiral samples. In addition to producing optically active chiral near fields, the metasurface also needs to provide a very high reflection coefficient (> 0.9999) with no polarization change at a high angle of incidence to maintain an EW-CRDS cavity. That means the loss due to the absorption and scattering of light by the metasurface should be kept very low such that the ring-down time is sufficiently long for the oscilloscope to resolve the time traces.

In the following, we discuss the performance of the metasurface as obtained from numerical simulations. The simulated reflectance spectrum of the metasurface at an incident angle of 70° is shown in Figure 2b. All the simulations are performed using the wave optics module in COMSOL 5.6 and further confirmed by Lumerical FDTD simulations. It can be noted from the figure that the reflectance of the metasurface depends on the wavelength of the incident light, and we observed an almost unity p-polarized reflection after a cut-off wavelength of 760 nm. For comparison, we also include the p-polarized reflection spectrum of a bare prism, a gold mirror of infinite thickness, and a gold metasurface of the same dimension as the dielectric one. The bare prism shows an ideal unity reflection coefficient due to TIR. However, the reflection coefficients of the gold metasurface and that of the gold mirror are significantly smaller than unity due to the absorption of light by gold. Although the metallic nanostructures are known to show a very high OC [12,13,23], the direct use in EW-CRDS is challenging because of the inherent absorption of light by the metal. In addition, for the gold metasurface, some distinct features around 750 nm are also observed, which can be associated with the surface lattice resonance found in periodic lattice structures [24,25].

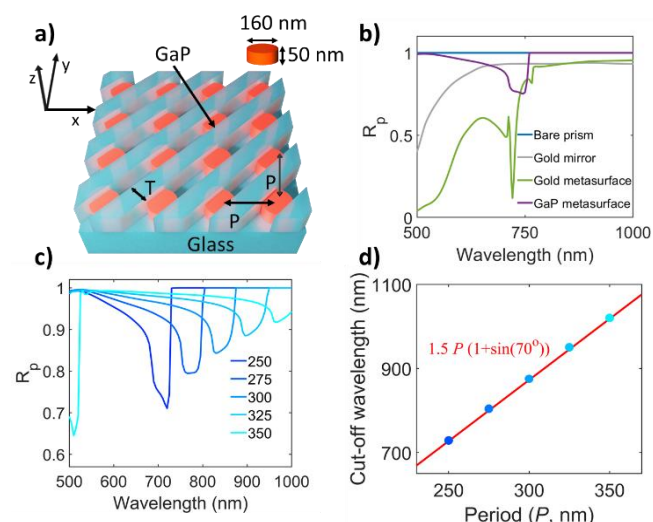

**Figure 2**: The dielectric metasurface and its optical response. (a) A small area of GaP metasurface (showing $4 \times 4$ unit cells) made of the GaP nanopillars arranged in a two-dimensional infinite array with a period of 260 nm on a glass substrate, a layer of glass stripes in the diagonal direction is used to partially cover the nanopillars. (b) The reflectance spectrum of the GaP metasurface (purple), a bare glass substrate (blue), a gold mirror (grey), and a gold metasurface (green) with the same dimension as the GaP metasurface with GaP nanopillars replaced by gold nanopillars. (c) The reflectance spectrum of GaP metasurface for different periods of the unit cell for a metasurface covered with water. (d) The cut-off wavelength for almost unit reflection coefficient is shown as a function of the lattice period, and the solid red line shows the cut-off obtained from the analytical expression.

The design parameters of the metasurface are also important for performance. Figure 2c shows the dependence of the reflection spectrum on the periodicity of the lattice, which indicates a periodicity-dependent cut-off wavelength for an almost unity reflection coefficient. The cut-off wavelength shows a direct proportionality with the periodicity of the grating, as shown in Figure 2d. The proportionality constant can be directly obtained from the condition of vanishing of higher diffraction order in the reflected light beam and no transmitted light in the surrounding medium [26]. Briefly, the condition for vanishing diffraction orders (m) can be obtained from the momentum-matching condition of the reflected ($k_x^R(m) = \frac{2\pi m}{d} + 1.5 \, 2\pi \sin 70°/\lambda$) and transmitted ($k_x^T(m) = 2\pi m/d + 1.5 \, 2\pi \sin 70°/\lambda$) diffraction orders. For non-zero transmission, $\lambda|m/P + 1.5 \sin 70°/\lambda|/1.33 \leq 1$, and, for non-zero higher diffraction orders ($|m| > 0$) in reflection, $|\lambda m/1.5P + \sin 70°| \leq 1$. For the geometry of the GaP metasurface, the condition for vanishing higher-order diffraction for the reflected beam also ensures a zero transmission, therefore, the cut-off wavelength of the metasurface is given as $\lambda = 1.5P(1 + \sin 70°)$. It is worth mentioning that in this calculation the polarization change caused by the glass stripes is neglected. Therefore, the polarization-preserving reflection coefficient of the metasurface is observed to be slightly less than unity even above this cut-off wavelength. The simple design of the metasurface allows us to reach an almost unit reflection coefficient with a finite average near field OC, which can be manipulated with the orientation of the structure relative to the incident light polarization. In principle, the broadband unit reflection provided by the proposed metasurface can be used to obtain the broadband CD response of the molecule in all the wavelengths above 800 nm using a tuneable pulse laser or perform broadband EW-CRDS on the system [27].

For the measurement of the CD response of the chiral entities, molecules can either be preconcentrated on the surface-functionalized nanostructure [28,29] or a flow channel can be designed to introduce the chiral molecules above the metasurface [30,31]. The absorption of the analyte on the surface of the nanostructure is detected by measuring the time-dependent intensity of light transmitted through the cavity mirror recorded by a light detector, such as a photomultiplier tube. The time trace of the transmission intensity can be analyzed to obtain the ring-down time ($\tau$). Assuming no other loss channels, the ring-down time of the system can be easily estimated using the following relation [19],

$$\tau = \frac{2l}{c \, ln(R_1 R_2 R_m^2)} \qquad (1)$$

Here, $c$ is the speed of light in air, $l$ is the distance between the two mirrors (assumed 1 m in the following), $R_1$ and $R_2$ are the reflection coefficient of the mirrors M1 and M2. $R_m$ represents the effective reflection coefficient of the metasurface with the chiral molecules. The adsorption of molecules to the metasurface or their presence in the near field of the metasurface changes the absorption of the medium surrounding the metasurface, which in turn changes the reflectance ($R_m$) of the metasurface and ultimately the ring-down time by a small amount $\Delta\tau$. The change in absorption for the two-handedness of the molecules (a measure of CD) is thus proportional to $\Delta\tau/\tau^2$.

In the following, we focus on the near-field OC distribution around the pillars and demonstrate that in-plane rotation would allow switching the handedness of the OC in the zone open to the analyte. Figure 3a shows the OC distribution of a unit cell under the excitation of the evanescent field of the p-polarized incident light (wavelength=800 nm, incident angle = 70°) undergoing TIR at the surface of the metasurface. Under linearly polarized excitation, the near-field OC exhibits a four-lobed pattern with alternating handedness around the pillar, i.e. the alternating blue-red lobes in Figure 3a. The diagonally orientated glass stripes (grey shaded areas in Figure 3a) cover the sites of near-field OC with one handedness and expose the other pair of sites with single-handed near-field OC. Therefore, analytes can only access the near-field OC around the pillar with one single handedness, making CD measurement possible. Note that the handedness of the OC does not vary within the height of the nanopillars (Figure 3b). Therefore, the in-plane glass stripes are sufficient. Also, the whole metasurface with the blocking glass stripes is immersed in water. Therefore, the index contrast at the boundaries of the glass stripes is reduced significantly. This greatly suppresses the influence of the glass stripes. To flip the handedness of the near-field OC accessible to the analyte, one just needs to rotate the structure in the plane by an azimuthal angle of 90° (Fig 3c). Since the rotation only changes the orientation of the blocking glass stripes to the opposite diagonal direction but does not vary the alternating four-lobed OC pattern near the pillar, the OC sites open to the analyte are switched to the other pair with the opposite OC. The possibility of flipping the handedness of OC of the open sites by a simple rotation makes it extremely useful for the experimental implementation as it allows a direct CD measurement from a chiral sample.

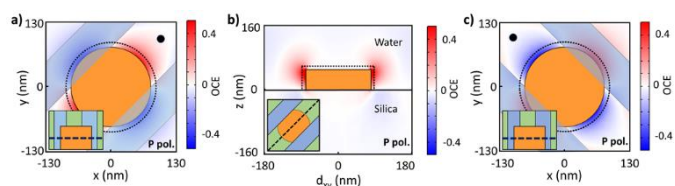

**Figure 3**: The spatial distribution (planes are shown in the inset) of the OC in the near field of nanostructure for a wavelength of 800 nm incident at an angle of 70° from the glass substrate. (a) The OC distribution in the XY plane for incident p-polarization. The black dashed line is used to mark the plotted plane on the structure and the green area is used to indicate the exposed area of the GaP metasurface. A dotted black curve is used to mark a distance of 10 nm from the surface of the nanoparticle that can be occupied/filled with a molecular layer. (b) The OC distribution in the Z plane along the diagonal of the unit cell ($d_{xy}$) of the metasurface for incident p-polarization. (c) The OC distribution in the XY plane for incident p-polarization with the same structure as in Fig. 3a, but rotated by an angle of 90°. The black dots in a) and c) indicate the in-plane orientation of the metasurface.

Finally, we evaluate the ring-down time and sensing sensitivity in the presence of chiral absorbers. A 10-nm layer of chiral molecule analyte of the various absorption coefficients was introduced to the vicinity of the metasurface. The molecular layer is modelled as an optical medium with a permittivity of $1.7689 + s\,10^{-2}i$ and Pasteur parameter $s(10^{-4} + 10^{-6}i)$ [32], where $s$ is a factor linked to the concentration of the molecules [33-36]. Since the relationship between $s$ and the concentration depends on the actual type of molecules, in this theoretical study, we choose to report the ring-down time with respect to $\log(s)$. The metasurface with a chiral molecular layer was modelled in COMSOL using the constitutive relations of an electromagnetic field in a chiral medium [32,37,38]. The reflection coefficient obtained from the simulation is inserted into equation 1 to calculate the ring-down time, assuming the reflection coefficient of the two mirrors ($R_1$ and $R_2$) to be 0.99995. Figure 4 shows the calculated difference in the ring-down time of the two enantiomers against the concentration factor of the Pasteur parameter $s(10^{-4} + 10^{-6}i)$, which representing the chiral optical constant of the molecular layer adsorbed on the exposed surface of GaP in the metasurface. For comparison, we also plot the calculated differences in the ring-down time for the same molecular film deposited on a GaP disc array without the blocking glass stripe arrays, on a surface with only the blocking glass stripe array without discs, and on the surface of a bare prism. Only the case with the GaP metasurface with correctly designed blocking glass strips shows a detectable difference in the ring-down time. The dramatic difference in the ring-down times obtained from the GaP disc array with and without the blocking glass stripe array shows the importance of having the blocking glass stripe array in the proposed GaP metasurface. Without the blocking glass stripe array, the sum of the optical chirality near the disc is almost zero due to the fact that the handedness of the optical chirality is alternating and thus cancel the optical chirality enhancement (Fig. 3). This results in no difference in the absorption and thus no difference in the ring-down time. In the cases of a bare prism surface and a surface with only blocking stripes, linearly polarised light does not generate any near field with significant optical chirality. Therefore, there is no difference in the ring-down time.

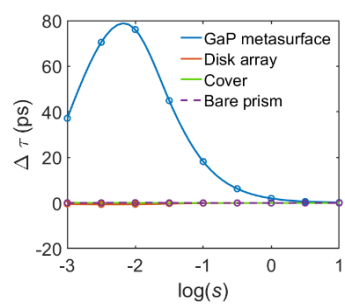

**Figure 4**: The difference in the ring-down time (Δτ) of the two enantiomers with respect to log (s), where s is the scaling factor of the Pasteur parameter $s(\mathbf{10^{-4} + 10^{-6}}i)$ that links the parameter to the sample concentration. The ring-down time differences for the EW-CRDS cavity using various metasurfaces are shown, including the GaP disc array with the blocking glass stripe array (blue trace), GaP disc array without the blocking glass stripe array, only the blocking glass stripe array without GaP discs (green trace) and the bare prism (dashed purple trace) for p-polarized excitation.

In the case of GaP disc array with the blocking glass stripe array, a large molecular absorption (high values of log(s)) leads to a very short ring-down time which results in negligible chiral absorption contribution amplification. On the other hand, for very small molecular absorption (low values of log(s)), the ring-down time increases significantly allowing maximal amplification of the chiral absorption contribution, but the Pasteur parameter ($s(10^{-4} + 10^{-6}i)$) becomes too small. The interplay of these two effects determines the maximum ring-down time difference, which indicates an optimum absorption of the molecule for best chiral sensitivity of the CRDS cavity with the proposed metasurface. The optimum sensitivity of the chirality is obtained at much lower absorption from the chiral molecules as compared to the previous reports [15,32], indicating the ability of the EW-CRDS to detect chirality at a much smaller concentration. For the case large absorption from the chiral molecules, the usual method of OC enhancement using plasmonic or dielectric nanostructure might be more suitable. Here, we note that the current metasurface is off resonance with the excitation laser. Therefore, the near optical field is rather weak and thus provides a small difference in the ring-down time in the range of 10 picoseconds. Although measuring such a difference in the ring-down time is possible [39], it requires a single mode excitation of a very stable optical cavity with low detection noise and fast oscilloscope. To enhance the ring-down time difference, the metasurface can be designed to exhibit resonances matching the excitation wavelength. However, this might also lead to larger absorption and scattering loss. Therefore, design strategies, such as inverse design using an evolutionary algorithm or rational design based on bound-state-in-the-continuum, can be explored to obtain a metasurface that provides high OC enhancement for linear polarized incidence, high reflection coefficient, and minimum polarization change.

## Conclusions

The integration of EW-CRDS with a dielectric metasurface giving an optically chiral near field provides an opportunity for ultrasensitive detection of the chiral molecules at a very low concentration. It can be used for detecting chiral molecules in a flow channel or in the thin film of the molecules deposited over the metasurface, with incident linear polarized light. Although all the studies are performed at the wavelength of 800 nm the metasurface provides almost unit reflection for the entire wavelength above 800 nm, making it a tool for broadband chiral detection. In this work, a small value of OC is obtained because of an off-resonance condition excitation of GaP in the metasurface. Nevertheless, the simple metasurface demonstrates the possibility to perform direct and ultrasensitive measurements of CD response from a chiral thin film in EW-CRDS using linearly polarized light.

## Author Contributions

JSH conceived the idea. AKS and ZHL performed the simulations. JSH, AKS and ZHL analysed the data. All authors participated in the discussions and contributed to the manuscript preparation.

## Conflicts of interest

There are no conflicts to declare.

## Acknowledgments

We thank the German science foundation for funding this research (Project numbers: 423427290, 445415315, and IRTG 2675-C1 "META-ACTIVE"). In addition, we acknowledge funding by the Free State of Thuringia under the number 2019 FGI 0028, co-financed by funds from the European Union under the European Regional Development Fund (ERDF). We also thank Dr. Ilya Shadrivov and Dr. Alexander V Kildishev for helpful discussions.